\documentclass[twocolumn,showpacs,preprintnumbers,amsmath,amssymb,superscriptaddress]{emulateapj}

\usepackage{graphicx}

\usepackage{aas_macros}

\usepackage{color}

\begin{document}


\title{A Case Against Spinning PAHs as the Source of the Anomalous
  Microwave Emission}%

\author{Brandon S. Hensley}
\email{brandon.s.hensley@jpl.nasa.gov}
\affiliation{Department of Astrophysical Sciences,  Princeton
  University, Princeton, NJ 08544, USA}
\affiliation{Jet Propulsion Laboratory, California Institute of Technology, 4800
Oak Grove Drive, Pasadena, CA 91109, USA}

\author{B. T. Draine}
\affiliation{Department of Astrophysical Sciences,  Princeton
  University, Princeton, NJ 08544, USA}

\author{Aaron M. Meisner}
\affiliation{Berkeley Center for Cosmological Physics}
\affiliation{Lawrence Berkeley
National Laboratory, Berkeley, CA 94720, USA}

\date{\today}

\begin{abstract}
We employ an all-sky map of the anomalous microwave emission (AME)
produced by component separation of the microwave sky to study correlations between the AME and Galactic dust
properties. We find that while the AME is highly correlated with all
tracers of dust emission, the best predictor of the AME
strength is the dust radiance. Fluctuations in the AME intensity per dust
radiance are uncorrelated with fluctuations in the emission from
polycyclic aromatic hydrocarbons (PAHs),
casting doubt on the association between AME and PAHs. The PAH
abundance is strongly correlated with the dust optical depth and dust
radiance, consistent with PAH destruction in low density regions. We find that the AME intensity increases
with increasing radiation field strength, at variance with predictions
from the spinning dust hypothesis. Finally, the temperature-dependence
of the AME per dust radiance disfavors the interpretation of the AME
as thermal emission. A reconsideration of
other AME carriers, such as ultrasmall silicates, and other emission
mechanisms, such as magnetic dipole emission, is
warranted.
\end{abstract}

\section{Introduction}
High sensitivity, full-sky observations of the far-infrared and
microwave sky from WMAP and {\it Planck} have pushed studies of the Cosmic Microwave Background
(CMB) to a regime in which contamination from Galactic foregrounds
has become a key uncertainty in the analysis. Understanding the
physical nature of the foreground
components and producing better models of each are essential for
mitigating this uncertainty.

The anomalous microwave emission (AME) is perhaps the least well-understood of the foreground
components. Discovered as a dust-correlated emission excess peaking
near 30\,GHz
\citep{Kogut+etal_1996, deOliveiraCosta+etal_1997,
    Leitch+etal_1997}, AME is often ascribed to electric dipole
emission from rapidly rotating ultrasmall dust grains
\citep{Draine+Lazarian_1998a, Draine+Lazarian_1998b, Hoang+Draine+Lazarian_2010,
  Ysard+Verstraete_2010, Silsbee+AliHaimoud+Hirata_2011},
i.e. ``spinning dust emission.'' Empirically, this emission component
peaks between $\sim20$ and 50\,GHz \citep[e.g.][]{Planck_Int_XV} and
has an emissivity per H of
$\sim3\times10^{-18}$\,Jy\,sr$^{-1}$\,cm$^2$\,H$^{-1}$ at 30\,GHz
\citep{Dobler+Draine+Finkbeiner_2009, Tibbs+etal_2010,
  Tibbs+etal_2011, Planck_Int_XV, Planck_Int_XVII}. Polycyclic
aromatic hydrocarbons (PAHs), which give rise to prominent
emission features in the infrared, are considered 
natural carriers of the AME due to their small size and apparent
abundance \citep{Draine+Lazarian_1998a}.

Theoretical spinning dust SED
templates based on a PAH size distribution that reproduces the
infrared emission features have been successful in fitting
observations of the AME both in the
Galaxy \citep{MivilleDeschenes+etal_2008, Hoang+Lazarian+Draine_2011,
  Planck_2015_X} and in the sole extragalactic AME detection in the
star-forming galaxy NGC\,6946
\citep{Murphy+etal_2010,Scaife+etal_2010,Hensley+Murphy+Staguhn_2015}. 

The AME has been observed to correlate well with the PAH emission
features in the infrared. \citet{Ysard+MivilleDeschenes+Verstraete_2010} found that,
over the full sky, the AME was more correlated with emission at 12\,$\mu$m
than with 100\,$\mu$m. Likewise, the AME in the dark cloud LDN 1622 was
better correlated with the 12 and 25\,$\mu$m emission than with either the 60
or 100\,$\mu$m emission \citep{Casassus+etal_2006}. However, studying
the AME in a sample of 98 Galactic clouds, \citet{Planck_Int_XV} found
no significant differences between the 12, 25, 60, and 100\,$\mu$m
emission in their correlation with the AME. Analysis of both the
Perseus molecular cloud \citep{Tibbs+etal_2011} and the H{\sc II}
region RCW175 found no compelling link between the PAH abundance and
the AME. Likewise, the link between the
AME and PAH abundance determined from dust model fitting has proven
tenuous in NGC\,6946 \citep{Hensley+Murphy+Staguhn_2015}.

In addition to spinning ultrasmall grains, grains containing ferro- or
ferrimagnetic materials are also predicted to radiate strongly in the microwave
and may contribute to the AME \citep{Draine+Lazarian_1999,
  Draine+Hensley_2013}. \citet{Draine+Lazarian_1999} argued that the
spinning dust and magnetic emission mechanisms could be distinguished by observing the AME in
dense regions where PAHs would likely be depleted due to
coagulation. If the AME per dust mass is constant across both dense
and diffuse regions, then spinning dust emission would be disfavored.

In this work, we test the spinning PAH hypothesis using new full-sky
observations of the infrared and microwave sky. \citet{Planck_2015_X} decomposed the {\it Planck}
sky into foreground components making use of both the 9-year WMAP data
and the Haslam 408 MHz survey. The combination of these data allowed the low
frequency foreground components -- primarily synchrotron, free-free,
and AME -- to be disentangled, producing a full-sky map of the AME.

All-sky WISE observations at $12\,\mu$m, a tracer
of PAH emission, have natural synergy with the AME map, allowing us to test at high significance
the link between the AME and PAHs. Additionally, all-sky dust modeling
by \citet{Planck_2013_XI} permits deeper exploration into the dust and
environmental parameters that influence the strength of the AME.

This paper is organized as follows: in Section~\ref{sec:data} we
summarize the data sets used in our analysis; in
Section~\ref{sec:tests} we describe the correlations predicted by our
current understanding of the spinning PAH hypothesis; in
Section~\ref{sec:analysis} we present the relationships between
environmental and dust properties and AME as derived from the data;
in Section~\ref{sec:discussion} we discuss the implications of these
results on spinning dust theory in particular and AME modeling in
general; and we summarize our principal conclusions in
Section~\ref{sec:conclusions}.

\section{Data}
\label{sec:data}
\subsection{{\it Planck} Foreground Separation Maps}
Combining full-mission all-sky {\it Planck} observations
\citep{Planck_2015_I} with the 9-yr. WMAP data \citep{Bennett+etal_2013} and the
Haslam 408 MHz survey
\citep{Haslam+etal_1982}, \citet{Planck_2015_X} used data
from 32 different detectors spanning a range of 408 MHz to 857 GHz in
frequency to perform foreground component separation within the Bayesian
\texttt{Commander} analysis framework \citep{Eriksen+etal_2004,Eriksen+etal_2006,Eriksen+etal_2008}. They
constructed theoretically-motivated models for the frequency
dependence of each component -- including the CMB, synchrotron, free-free, thermal
dust, and AME -- while minimizing the number of free
parameters needed and simultaneously fitting for calibration offsets. Using a Gibbs sampling algorithm, they
determined the best-fit values
for each model parameter on a
pixel-by-pixel basis.

In this work, we focus primarily upon the resulting map of the
AME. This component was modeled by the sum of {\it two} spinning dust
spectra with fixed spectral shape as determined by the \texttt{SpDust}
code \citep{Ali-Haimoud+Hirata+Dickinson_2009,
  Silsbee+AliHaimoud+Hirata_2011}, but differing amplitudes and peak
frequencies. One of the spectra was required to have a spatially fixed
peak frequency, fit to be 33.35 GHz, while the other peak frequency was allowed to freely vary from pixel
to pixel. Thus, the data products consist of an amplitude for each
AME component, the peak frequency of the spatially varying component,
and the uncertainty of each for every pixel on the sky.

To facilitate comparisons with the literature, we calculate the sum of
the two components at 30\,GHz. We also calculate the uncertainty in
this quantity assuming Gaussian errors and ignoring the uncertainty on
the peak frequency of both components.

In addition to the AME map, we employ the parameters from the
thermal dust fit to compute the 353\,GHz dust optical depth
$\tau_{353}$. Although this parameter was also an explicit data product of
the full-sky modified blackbody fits performed by \citet{Planck_2013_XI},
we prefer to use the results from \citet{Planck_2015_X} for several
reasons. First, the fits were performed on more data and with more
detailed treatment of calibration and bandpass uncertainties. Second,
the fits were performed at $1^\circ$ versus $5'$ resolution,
mitigating the effects of cosmic infrared background (CIB)
anisotropies. Finally, a single-temperature 
modified blackbody model has been
demonstrated to be inadequate to fit the FIR dust emission from {\it Planck} HFI
frequencies to 100\,$\mu$m \citep{Meisner+Finkbeiner_2015, Planck_2015_X}. Thus, we prefer
to use a $\tau_{353}$ derived from fits to the Rayleigh-Jeans portion
of the dust emission spectrum only. We find $\tau_{353}$ derived in
this way to be on average 10\% lower than that reported by
\cite{Planck_2013_XI}.

Additionally we employ the parameter maps that characterize the free-free,
synchrotron, and CO emission. As free-free and CO emission cannot be fit
reliably in regions of low surface brightness, we exclude all pixels
with emission measure of 0.1\,cm$^{-6}$\,pc or less and with no fit
CO emission when analyzing the former and latter, respectively.

The \texttt{Commander} parameter maps have a resolution of
$1^\circ$ (FWHM) and are pixellated with \texttt{HEALPix} \citep{Gorski+etal_2005} resolution of ${\rm
  N}_{\rm side} = 256$. For our analysis, we downgrade these maps to
$N_{\rm side} = 128$, corresponding to pixels of about 27'.5 on a side.

\subsection{Dust Parameter Maps from Modified Blackbody Fitting}
A key limitation of the thermal dust fits performed by
\citet{Planck_2015_X} is the omission of any data at higher
frequency than the {\it Planck} 857\,GHz band. Without information on
the Wien side of the dust emission spectrum, it is difficult to
constrain the dust temperature and luminosity.

Thus, we employ the full-sky parameter maps of \citet{Planck_2013_XI}, who fit a
modified blackbody model of the dust emission to the 2013 {\it Planck}
857, 545, and 353\,GHz data as well as 100\,$\mu$m data from IRAS. For
the latter, they employ both the reprocessed IRIS map
\citep{MivilleDeschenes+Lagache_2005} as well as the map of \citet{Schlegel+Finkbeiner+Davis_1998}. The fits were
performed at $5'$ resolution and yielded an estimate of $\tau_{353}$,
the dust radiance $\mathcal{R} \equiv \int I_\nu^{\rm dust}{\rm d}\nu$, the dust temperature $T_d$, and the
dust spectral index $\beta \equiv {\rm d\ ln}\,\tau/{\rm d\ ln}\,\nu$ for each pixel. We
employ $\mathcal{R}$, $T_d$, and $\beta$ from these fits in our
analysis, but use $\tau_{353}$ from the \texttt{Commander}
fits as discussed above.

We use the parameter map pixellated with \texttt{HEALPix} resolution of ${\rm
  N}_{\rm side} = 2048$, then smooth with a Gaussian beam of FWHM
$1^\circ$  and downgrade the resolution to ${\rm
  N}_{\rm side} = 128$.

\subsection{WISE 12\,$\mu$m Map}
\label{subsec:data_wise}
WISE observed the full sky in four infrared bands -- 3.4, 4.6, 12, and
22\,$\mu$m \citep{Wright+etal_2010}. The 12\,$\mu$m channel ($W3$),
sensitive from 7 to 17\,$\mu$m,
captures the strongest of the infrared emission features
associated with PAHs and thus traces the population of small dust
grains. The full-sky WISE 12\,$\mu$m imaging has been reprocessed by
\citet{Meisner+Finkbeiner_2014} to isolate diffuse emission from Galactic 
dust at 15'' resolution.

Because interplanetary dust models are insufficiently accurate to 
remove zodiacal light from the $W3$ data,
\citet{Meisner+Finkbeiner_2014} adopted a zero level based on the
$Planck$ 857\,GHz map, essentially replacing 12\,$\mu$m modes on scales
larger than 
2$^{\circ}$ with a rescaling of 857\,GHz dust emission. Since we wish to study
real fluctuations in PAH emission per FIR dust radiance on $1^{\circ}$ 
scales, this zero level procedure has the potential to artificially suppress
part of our signal.

Our analysis relies on quantifying the PAH abundance in a given pixel
through the ratio of the $W3$ emission to the FIR dust radiance
$\mathcal{R}$, a quantity we designate $f_{\rm PAH}$ (see
Section~\ref{subsec:model}). Computing $f_{\rm PAH}$ via simple division of the
\citet{Meisner+Finkbeiner_2014} $W3$ map by $\mathcal{R}$ is problematic because this
method is sensitive to the $Planck$ 857\,GHz-based 12\,$\mu$m large-scale zero
level.

We therefore employ an alternative approach to compute $f_{\rm PAH}$,
leveraging the excellent angular resolution of WISE and $Planck$ to
render the large-scale zero level of the $W3$ map irrelevant. Our
principal analysis employs $N_{\rm side} = 128$ \texttt{HealPix}
pixels of $\sim$0.5$^{\circ}$ on a side. To determine $f_{\rm PAH}$,
we subdivide each of these pixels into approximately 120 subpixels of
$2.5'$ on a side and then derive the best-fit correlation slope of 
$W3$ emission versus $\mathcal{R}$ across the 120 subpixels. To infer
the optimal $f_{\rm PAH}$ value within a single \texttt{HEALPix} pixel, we
employ the model

\begin{equation}
\label{eq:fpah}
\left(\nu I_{\nu}\right)_i^{12_{\mu m}} = \mathcal{R}_i \times f_{\rm PAH} + C
~~~,
\end{equation}
where, in the $i$th subpixel, $(\nu I_{\nu})_i^{12_{\mu m}}$ is the $W3$ map smoothed to $5'$
resolution and high-pass filtered at $15'$, and $\mathcal{R}_i$ is the
$Planck$ radiance map \citep{Planck_2013_XI} high-pass filtered at $15'$.
The high-pass filtering serves to remove large-scale modes from the comparison,
and the offset $C$ is a nuisance parameter. Because linear regression can become sensitive 
to the choice of fitting methodology in the limit of poor
signal-to-noise \citep[see, e.g.,][]{Hogg+Bovy+Lang_2010}, we restrict
analyses of these $f_{\rm PAH}$ 
correlation slopes to \texttt{HEALPix} pixels with very strong linear correlation 
between $(\nu I_{\nu})^{12_{\mu m}}$ and $\mathcal{R}$, using a 
threshold of Pearson $r$ $>$ 0.6. This threshold roughly corresponds to a 10\% 
fractional uncertainty on the inferred $f_{\rm PAH}$.

The $f_{\rm PAH}$ map derived in this way is qualitatively similar to that 
obtained by simply dividing the 12\,$\mu$m map by $\mathcal{R}$, but displays 
a factor of $\sim$2 broader range of $f_{PAH}$ values. Our basic conclusions
are unaffected by the choice of $f_{\rm PAH}$ map, and we employ the map obtained 
with Equation~\ref{eq:fpah} as the default.

Only in the case of Figure~\ref{fig:wise_corr} do we require a map of the 12\,$\mu$m emission itself,
rather than $f_{\rm PAH}$. For this purpose, we generated a custom version
of the \citet{Meisner+Finkbeiner_2014} $W3$ map by replacing only modes larger
than 4.5$^{\circ}$ with a rescaling based on the $Planck$ 857\,GHz
emission. This better preserves real fluctuations in $12\,\mu$m emission
per $\mathcal{R}$ on $\simeq1^\circ$ scales.

Both our final maps of WISE 12\,$\mu$m and $f_{\rm PAH}$ are
pixellated with \texttt{HEALPix} resolution of ${\rm
  N}_{\rm side} = 128$.

\subsection{IRAS 100\,$\mu$m Map}
As our analysis is performed on scales of $\simeq 1^\circ$, we follow
\citet{Planck_2013_XI} in employing the IRAS 100\,$\mu$m map of
\citet{Schlegel+Finkbeiner+Davis_1998}. We do not however employ the
reprocessed IRIS maps \citep{MivilleDeschenes+Lagache_2005} on small
scales as was done in \citet{Planck_2013_XI}. We note that using the
IRIS map instead in our analysis somewhat degrades the tightness of the
correlation between the 100\,$\mu$m emission and the AME, particularly
in regions of low surface brightness. The effect is small and has
limited impact on our conclusions. 

We smooth the IRAS 100\,$\mu$m map of
\citet{Schlegel+Finkbeiner+Davis_1998} to a FWHM of $1^\circ$ and
pixellate with ${\rm N}_{\rm side} = 128$.

\subsection{Masks}
\label{subsec:galmask}

\begin{figure*}
    \centering
        \scalebox{0.8}{\includegraphics{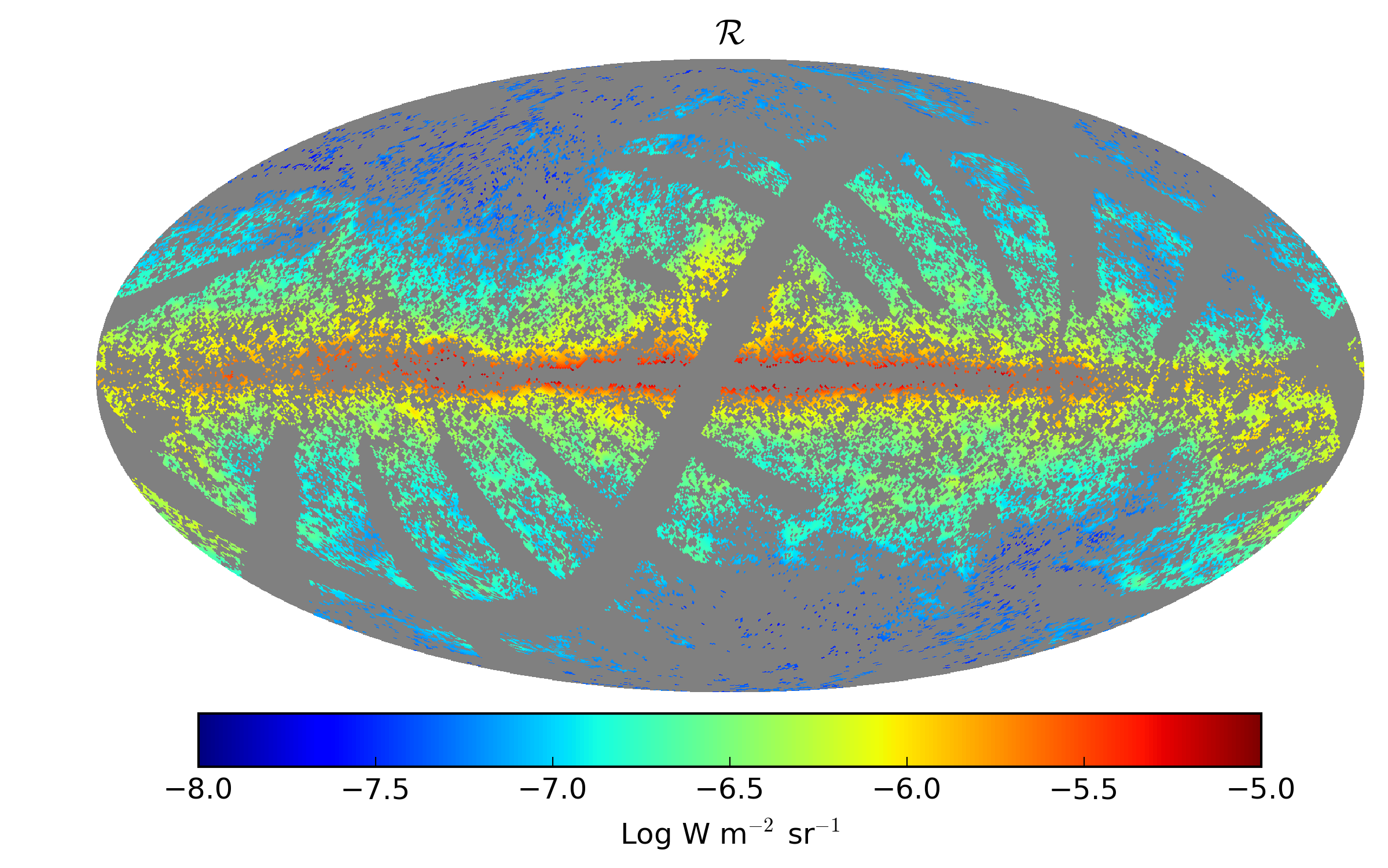}}  
        \scalebox{0.8}{\includegraphics{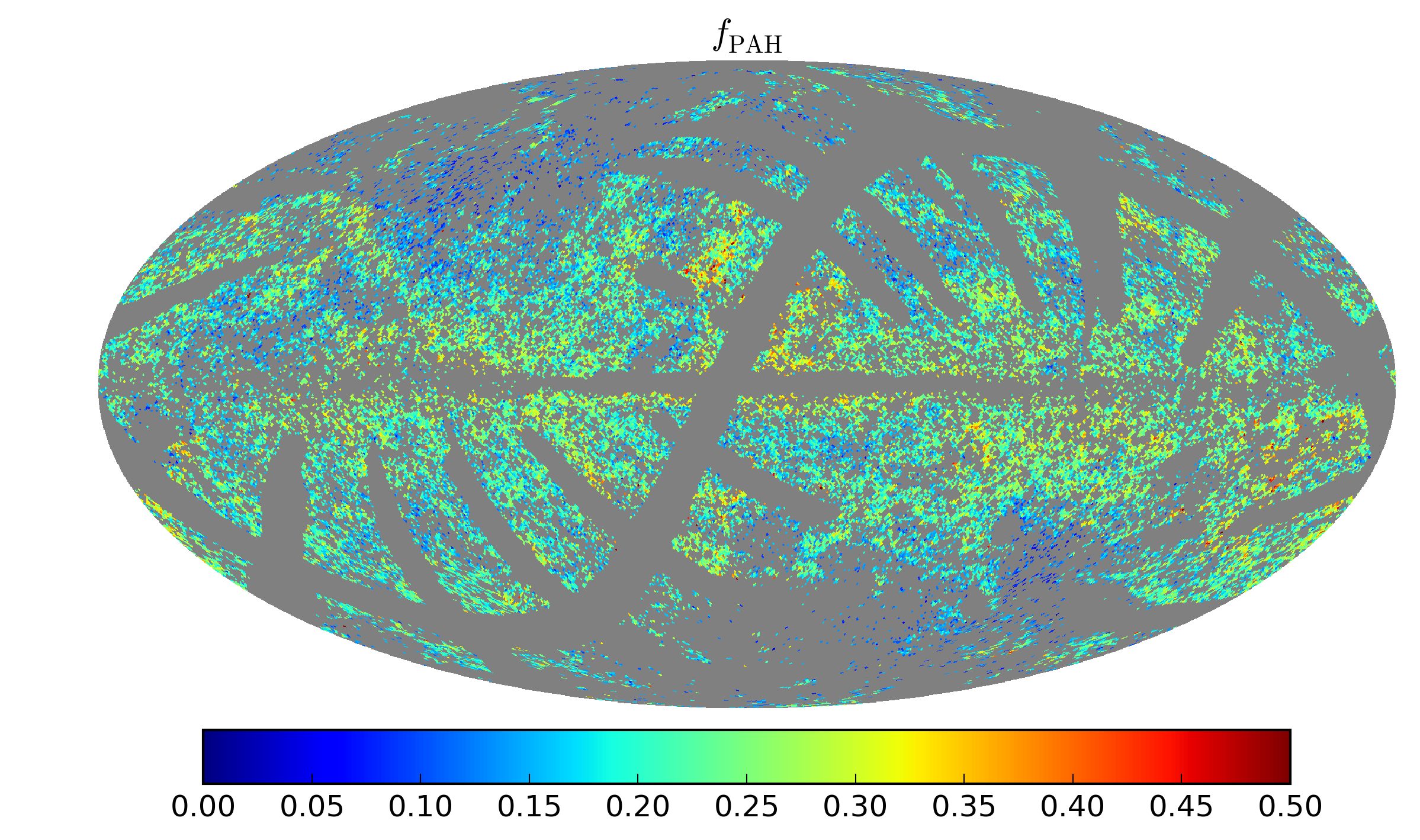}}
    \caption[Dust Radiance with Mask]{{\it Top}: The Mollweide projection full-sky map of $\mathcal{R}$ as derived by
      \citet{Planck_2013_XI} with the mask described in
      Section~\ref{subsec:galmask} overlaid in gray. The unmasked area
    comprises 26\% of the sky. {\it Bottom}: The full-sky map of
    $f_{\rm PAH} \equiv \left(\nu I_\nu\right)^{12\,\mu{\rm
        m}}/\mathcal{R}$ with the same mask overlaid.} \label{fig:mask_r} 
\end{figure*}

Zodiacal light in the ecliptic plane can dominate the signal from PAH emission at
12\,$\mu$m even on small angular scales, thereby biasing our primary PAH tracer. We therefore
exclude all pixels within five degrees of the ecliptic plane (8.7\% of
the sky) to mitigate this effect.

Likewise, artifacts from moon contamination are also present in the
12\,$\mu$m map. We therefore mask all pixels flagged for moon
contamination by \citet{Meisner+Finkbeiner_2014}, totaling 16\%
of the sky. Since our analysis requires that both $\mathcal{R}$ and
$I_\nu^{12\,\mu{\rm m}}$
measure true dust emission, and since these quantities will closely
correlate when they do, we require every pixel in our analysis to have a Pearson $r >
0.6$ between $\mathcal{R}$ and $I_\nu^{12\,\mu{\rm m}}$ across its 2.5'
subpixels as described in Section~\ref{subsec:data_wise}. This cut alone eliminates 42\% of the sky, mostly at low
surface brightness and high Galactic latitude where WISE lacks
sufficient sensitivity to measure diffuse dust emission.

Though nominally a full-sky mission, IRAS did not have 100\% sky
coverage. We therefore mask the regions IRAS did not observe
\citep{Wheelock+etal_1994}, totaling
2\% of the sky, as the 100\,$\mu$m emission in these regions estimated
using lower resolution DIRBE data \citep{Schlegel+Finkbeiner+Davis_1998}
may not be reliable.

To mitigate the effects of point sources, we employ the {\it Planck}
point source masks in intensity at each
HFI\footnote{\url{http://irsa.ipac.caltech.edu/data/Planck/release\_2/ancillary-data/masks/HFI\_Mask\_PointSrc\_2048\_R2.00.fits}}
and
LFI\footnote{\url{http://irsa.ipac.caltech.edu/data/Planck/release\_2/ancillary-data/masks/LFI\_Mask\_PointSrc\_2048\_R2.00.fits}}
band \citep{Planck_2015_I}. These masks eliminate
point sources with signal-to-noise greater than 5 at resolution
$N_{\rm side} = 2048$. We downgrade these masks to $N_{\rm side} =
128$ by rejecting any pixel containing a point source, resulting in
39\% of the sky being masked due to point sources.

Finally, we use the {\it Planck} Galactic plane mask covering 1\% of
the sky based on the 353\,GHz HFI
data\footnote{\url{http://irsa.ipac.caltech.edu/data/Planck/release\_2/ancillary-data/masks/HFI\_Mask\_GalPlane-apo0\_2048\_R2.00.fits}}
\citep{Planck_2015_I}. This eliminates the regions in the Galactic plane
with the most complicated and highest intensity emission and where the
relatively simple foreground models are most likely to break
down.

After applying the masks described above, 51,579 pixels covering 26\% of the sky
remain. The total mask is illustrated in Figure~\ref{fig:mask_r} on the
full-sky $\mathcal{R}$ map. This combination of masks is used in all analysis, although
we discuss the sensitivity of our results to various additional
masking (e.g. masking the Galactic plane) in
Section~\ref{subsec:masks}.

\section{Tests of AME Theory}
\label{sec:tests}
\subsection{AME Theory}
\label{sec:theory}
The AME has been suggested to be electric dipole emission from ultrasmall,
rapidly-rotating grains that have been torqued up through interactions
with both the gas and radiation field \citep{Draine+Lazarian_1998b,
  Hoang+Draine+Lazarian_2010, Ysard+Verstraete_2010,
  Silsbee+AliHaimoud+Hirata_2011}. The observed peak frequency of the
emission requires that the grains be small (radius $a \lesssim 10$\,\AA),
leading to a natural association with the PAHs that produce emission
features in the infrared.

If so, we might expect a linear relation between the
total PAH surface density $\Sigma_{\rm PAH}$ and the AME intensity. Adopting an empirical
30\,GHz spinning dust
emissivity of $3\times10^{-18}$\,Jy\,sr$^{-1}$\,cm$^2$\,H$^{-1}$, and
assuming this corresponds to typical Galactic values of $M_d/M_H =
0.01$ and $\Sigma_{\rm PAH}/\Sigma_d = 0.046$ \citep{Draine+Li_2007}, we would expect

\begin{equation}
I_{\nu, 30\,{\rm GHz}}^{\rm AME} = 0.8 \left(\frac{\Sigma_{\rm
      PAH}}{{\rm M}_\odot\,{\rm kpc}^{-2}}\right)\,{\rm Jy\, sr}^{-1}
~~~.
\end{equation}

In the context of the spinning dust model, environmental factors can
influence both the peak frequency of the emission and the emissivity
itself. The importance of collisions with ions depends on the
fractional ionization and the charge state of the ultrasmall
grains. In regions with very intense radiation fields,
drag forces from IR photon emission become important. Damping by the
rotational emission generally causes the
ultrasmall grains to have sub-LTE rotation rates.

The electric dipole moment distribution of the dust population will also
strongly influence the emissivity, though we have no {\it a priori}
estimates of the systematic variations of this quantity from one
region to another.

We expect spinning dust emission to be relatively insensitive to the
strength of the radiation field, which is an important source of
excitation only in fairly extreme environments such as
reflection nebulae and photo-dissociation regions (PDRs). However, the
emissivity per unit gas mass should increase with increasing local gas density, which
may correlate with the radiation field strength.

The impact of these factors on the 30\,GHz AME flux density was estimated
by \citet{Draine+Lazarian_1998b} to be only about a factor of two
between the Cold Neutral Medium, Warm Neutral Medium, and Warm Ionized
Medium environments. In our study, these effects are mitigated by the low angular
resolution of the maps, which are likely to be sampling dust emission
from different environments within each pixel.

These caveats notwithstanding, the spinning
PAH model for the AME predicts:

\begin{enumerate}
\item A linear correlation of the PAH surface density and the AME flux
  density at 30 GHz.
\item Relatively constant AME per PAH surface density over a range of
  radiation field strengths.
\end{enumerate}

\subsection{Data Model}
\label{subsec:model}
$\tau_{353}$ is equal to the product of the dust mass column density
and the dust opacity at 353\,GHz. Because dust grains are much smaller
than the wavelength of light at this frequency, the dust opacity is
insensitive to the size distribution of the grains and $\tau_{353}$ is
thus a reliable indicator of the total dust column density. There is
evidence, however, that the dust optical properties change somewhat in
different environments \citep{Planck_Int_XXIX}.

The dust radiance $\mathcal{R}$ is the frequency-integrated dust
intensity. $\mathcal{R}$ is estimated using the best-fit single temperature modified blackbody with
$\tau_\nu \propto \nu^\beta$, and is a tracer of both the dust column density and
the strength of the radiation field heating the dust.

Finally, the 12\,$\mu$m flux density is effectively a measure of the
total power emitted by PAHs, i.e., the starlight power absorbed by PAHs. Thus, the ratio of the 12\,$\mu$m
emission to the dust radiance is a proxy for the fraction of dust in
PAHs (see Equation~\ref{eq:fpah}). The product $f_{\rm PAH}\tau_{353}$ is then a proxy for the PAH column density. It is this
quantity which the spinning dust model
predicts to be the most accurate predictor of the strength of the
AME.

We assume that $\tau_{353}$, $f_{\rm PAH}\tau_{353}$, $\mathcal{R}$,
$I_\nu^{12\,\mu{\rm m}}$, and $I_\nu^{100\,\mu{\rm m}}$ correlate
with the AME intensity in a linear way, i.e., for each pixel $i$ and
each observable $A_i$,
the AME intensity in that pixel is given simply by $\alpha_i A_i$ where $\alpha_i$ is a
constant to be determined.

To identify the physical quantity
which is the best predictor of the AME intensity, we wish to quantify
the intrinsic dispersion in $\alpha_i$ across all pixels. We assume that each pixel
samples a Gaussian distribution with mean $\alpha$
and standard deviation $\sigma_\alpha$.

The likelihood of this model
given the data over all pixels is:

\begin{equation}
\label{eq:likelihood}
\mathcal{L} \propto \prod\limits_i \frac{1}{\sqrt{\sigma_{I,i}^2 +
    \left(A_i\sigma_\alpha\right)^2 }}\times{\rm exp}\left[\frac{-\left(I_{\nu,\ 30\,{\rm GHz},i}^{\rm AME} -
    \alpha A_i\right)^2}{\sigma_{I,i}^2 +
  \left(A_i\sigma_\alpha\right)^2}\right]
~~~,
\end{equation}
where $\sigma_{I,i}$ is the uncertainty on $I_{\nu,\ 30\,{\rm GHz},i}^{\rm
  AME} $. We thus seek the maximum likelihood values of $\alpha$ and
$\sigma_\alpha$ for each observable $A$. In practice, we employ the
\texttt{emcee} Markov chain Monte Carlo code \citep{Foreman-Mackey+etal_2013} to derive the best fit
values and confidence intervals for each parameter assuming uninformative
priors.

If there are zero point errors in any of the maps used in this
analysis, then $I_{\nu,\ 30\,{\rm GHz}}^{\rm AME} = \alpha A + b$ is
a more appropriate functional form for the relationship, where $b$ is
a parameter to be fit. We find that the introduction of an intercept
has little impact on the results, such as the ordering of the
$\sigma_\alpha/\alpha$, or the value of $\alpha$.

\section{Results}
\label{sec:analysis}
\subsection{Correlation with PAH Abundance}

\begin{figure*}
    \centering
        \scalebox{1.0}{\includegraphics{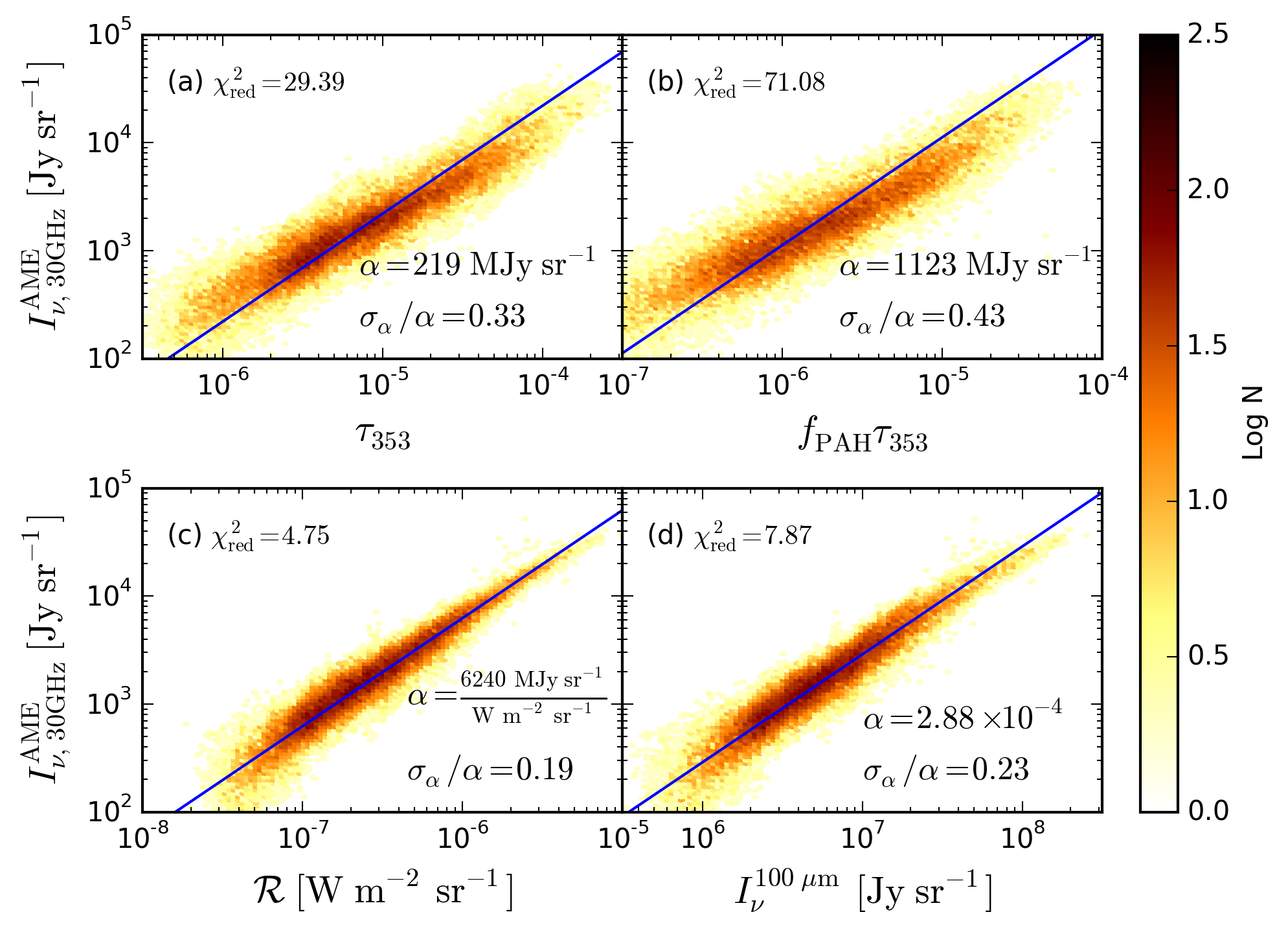}}  
    \caption[Correlations Between $I_\nu^{\rm AME}$ and Dust Observables]{The 30\,GHz AME intensity is plotted against (a)
      $\tau_{353}$, (b) $f_{\rm PAH}\tau_{353}$, (c) $\mathcal{R}$,
      and (d) $I_\nu^{100\,\mu{\rm m}}$. We divide the plot area into
      hexagons of equal area in log-space and color each according to the number of
      pixels that fall within that hexagon. Each panel has the same
      logarithmic area allowing for straightforward comparisons
      between panels. In doing this, we have restricted the range of
      each plot to exclude some outlying points. In each panel we also
    plot (solid blue) the line with slope equal to the best-fit value
    of $\alpha$. All four
    quantities are excellent tracers of the AME, but it is clear that
    $\mathcal{R}$ traces the AME with the greatest fidelity and
    least dispersion.} \label{fig:corr} 
\end{figure*}

\begin{deluxetable}{cccc}
      \tablewidth{0pc}
      \tablecaption{Correlation Analysis \\ $I_{\nu,\ 30\,{\rm GHz}}^{\rm AME}
      = \alpha A$\label{table:corr}}
    \tablehead{\colhead{$A$} & \colhead{$\alpha\pm\sigma_\alpha$} & \colhead{$\sigma_\alpha/\alpha$} & \colhead{$\sigma_A/A\left(0.2\right)$}}
    \startdata
    $\mathcal{R}$ & $\frac{6240\pm 1210}{\rm W\,m^{-2}\,sr^{-1}}$ MJy\,sr$^{-1}$ & 0.19 & 0.01 \\
    $I_\nu^{12\,\mu{\rm m}}$ & $(  7.48\pm   1.66)\times10^{-3}$  & 0.22 & 0.09 \\
    $I_\nu^{100\,\mu{\rm m}}$ & $(  2.88\pm  0.65)\times10^{-4}$ & 0.23 & 0.10 \\
    $\tau_{353}$ & 219 $\pm$  72\,MJy\,sr$^{-1}$ & 0.33 & 0.27 \\
    $f_{\rm PAH}\tau_{353}$ & 1123 $\pm$ 487\,MJy\,sr$^{-1}$ & 0.43 & 0.39
    \enddata
    \tablecomments{$\alpha$ and $\sigma_\alpha$ are the best fit
    values obtained from Equation~\ref{eq:likelihood} for each
    observable $A$ such that
    $I_{\nu,\ 30\,{\rm GHz}}^{\rm AME} = \alpha A$. The formal uncertainties
    on $\alpha$ and $\sigma_\alpha$ are less than the quoted accuracy
    in the table, generally of order 0.1\%. $\sigma_A/A\left(0.2\right)$ denotes the value
    of $\sigma_A/A$ such that $\sigma_\alpha/\alpha = 0.2$, indicating the level of
    observational uncertainty needed in $A$ to produce a relationship with $I_{\nu,\ 30\,{\rm GHz}}^{\rm AME}$
    comparably tight as that observed with $\mathcal{R}$.}
\end{deluxetable}

We test the predictions of the spinning PAH hypothesis laid out in
Section~\ref{sec:theory} by relating the observational data to
physical properties of the dust through the model described in
Section~\ref{subsec:model}.

In Figure~\ref{fig:corr} we plot the AME flux density at 30 GHz
against $\tau_{353}$, $f_{\rm PAH}\tau_{353}$, the dust radiance $\mathcal{R}$,
and the 100\,$\mu$m flux
density $I_\nu^{100\,\mu{\rm m}}$. All four correlate highly with
AME as expected, and we present the fit slope of the relation with
each in Table~\ref{table:corr}. The tightness of each correlation is
indicated by $\sigma_\alpha/\alpha$, with
$\mathcal{R}$ having the tightest correlation (see
Figure~\ref{fig:corr}c).

If the AME arises from spinning PAHs, we would expect the emission to
correlate better with $f_{\rm PAH}\tau_{353}$ than $\tau_{353}$. While
it is clear that both are good tracers of AME, $f_{\rm
  PAH}\tau_{353}$ has a {\it larger} dispersion about the best-fit
relation than $\tau_{353}$. Maximizing
Equation~\ref{eq:likelihood}, $A=\tau_{353}$ yields a relation with
$\sigma_\alpha/\alpha = 0.33$ while $A=f_{\rm PAH}\tau_{353}$ yields
$\sigma_\alpha/\alpha = 0.43$ (see Table~\ref{table:corr}). Thus,
$f_{\rm PAH}$ does not appear to contain additional information about
the strength of the AME not already present in $\tau_{353}$.

The spinning PAH model predicts that variations in the AME intensity
per unit dust mass should arise from variations in the
abundance of small grains. Therefore, as a second test of the link
between the AME and PAHs, we look for correlations between $f_{\rm
  PAH}$ and the AME per $\tau_{353}$. 

\begin{figure*}
    \centering
        \scalebox{1.0}{\includegraphics{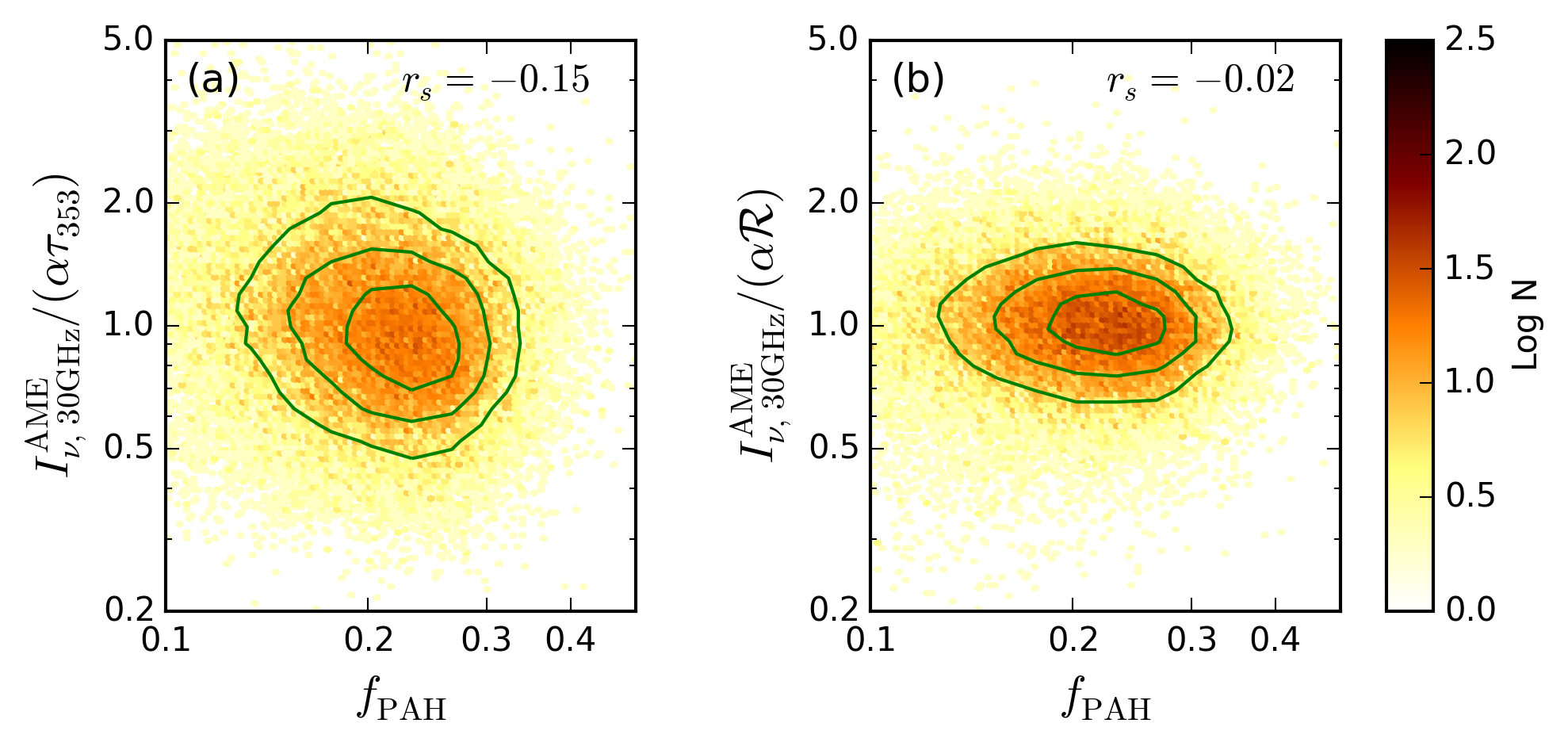}}  
    \caption[Non-correlation Between $f_{\rm PAH}$ and the AME]{$f_{\rm PAH}$ is plotted against the AME intensity
      normalized by (a) $\alpha\tau_{353}$ and (b)
      $\alpha\mathcal{R}$. Here and in subsequent plots, the
      isodensity contours corresponding to 25, 50, and 75\% of the
      pixels enclosed are plotted in green. There
      is no apparent correlation between $f_{\rm PAH}$ and the AME, at
      variance with the spinning PAH hypothesis. } \label{fig:residual} 
\end{figure*}

In Figure~\ref{fig:residual}a we
plot the AME intensity normalized by $\alpha\tau_{353}$ against
$f_{\rm PAH}$, but we find no evidence for the expected positive
correlation. We quantify this with the Spearman rank
correlation coefficient $r_s$ which, unlike the Pearson correlation
coefficient, does not assume a functional form for the relationship
between the two variables. We find $r_s = -0.15$, suggesting {\it
  anti}-correlation. 

Similarly, quantifying the correlation
between $f_{\rm PAH}$ and the AME intensity normalized instead by
$\alpha\mathcal{R}$ (see Figure~\ref{fig:residual}b) yields $r_s =
-0.02$, suggesting that $f_{\rm PAH}$ carries no information on the
AME intensity not already present in the dust radiance.

\begin{figure*}
    \centering
        \scalebox{1.0}{\includegraphics{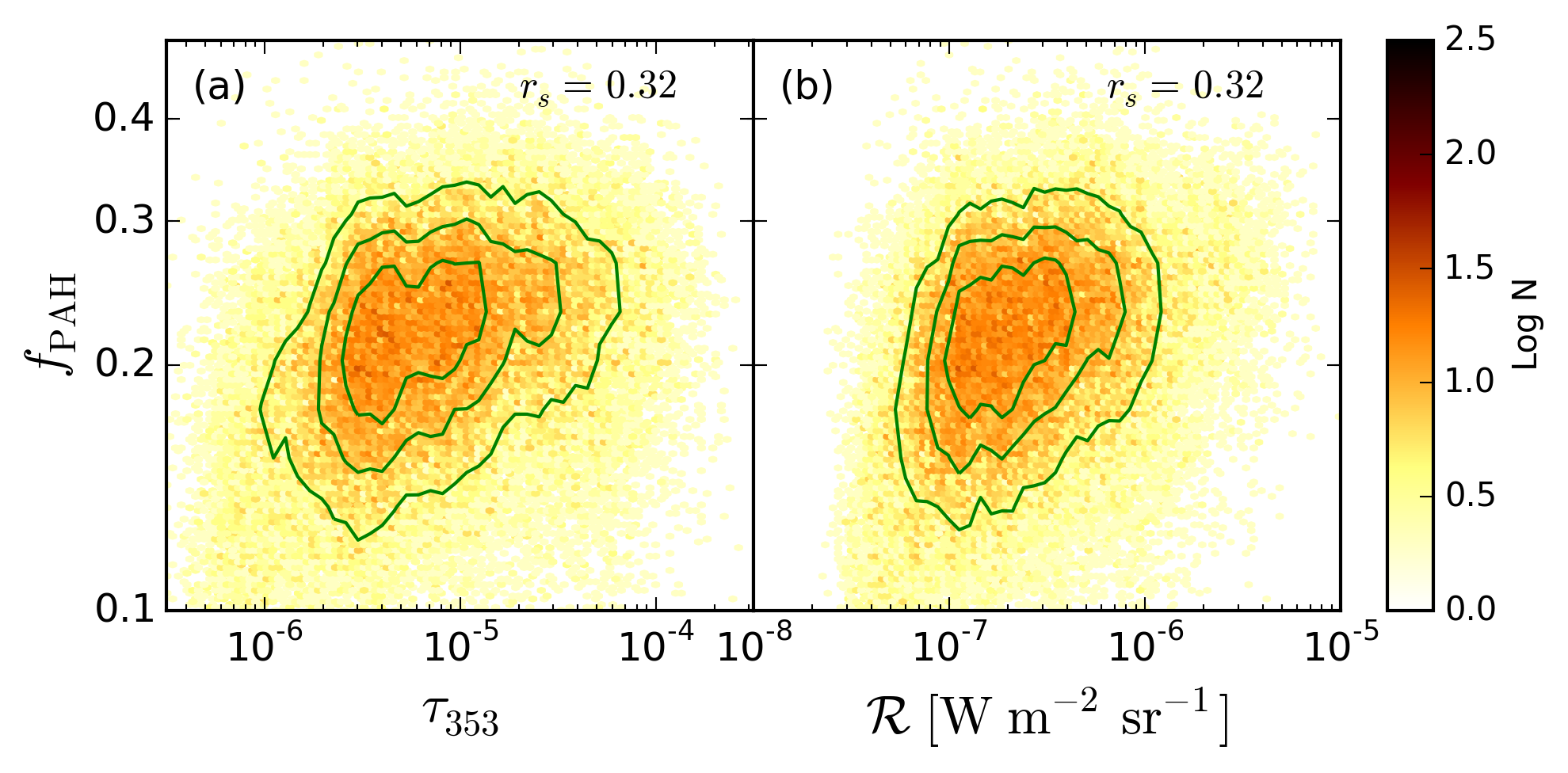}}  
    \caption[Correlation Between $f_{\rm PAH}$ and Both $\tau_{353}$
    and $\mathcal{R}$]{Plotting $f_{\rm PAH}$ against both
      (a) $\tau_{353}$ and (b) $\mathcal{R}$, it is clear that a strong
      positive correlation is present with both, consistent with
      depletion of PAHs in the diffuse ISM. The majority of the
      pixels have $0.15 < f_{\rm PAH} < 0.30$, suggesting that
      fluctuations in PAH abundance could account for a scatter of a
      factor of at most $\sim1.5$ in AME intensity unit per dust mass.} \label{fig:fpah_tau_r}
\end{figure*}

We note that $f_{\rm PAH}$ is itself correlated with both $\tau_{353}$
and $\mathcal{R}$ as demonstrated in
Figure~\ref{fig:fpah_tau_r}. $f_{\rm PAH}$ (measured on $\sim 1^\circ$
scales) varies from $\sim0.15 - 0.30$ over most of the unmasked sky, so that variations in
the PAH abundance might have been expected to account for a factor of $\sim1.5$
dispersion in AME intensity per dust mass. It is therefore striking
that we detect no correlation of AME/$\mathcal{R}$ with $f_{\rm
  PAH}$. 

The observed strong
correlation between $f_{\rm PAH}$ and $\tau_{353}$
is evidence for PAH destruction
in the diffuse ISM (low $\tau_{353}$). Depletion of PAHs by a
factor of $\sim3$ to suppress photoelectric heating in the warm ionized medium
(WIM) has been invoked to explain high ratios of H$\alpha$ to
free-free emission \citep{Dong+Draine_2011} and is roughly consistent with the
range of $f_{\rm PAH}$ we observe.

A study of the H$\alpha$-correlated AME by
\citet{Dobler+Draine+Finkbeiner_2009} found the AME to be a factor of
$\sim3$ less intense than the amplitude calculated by
\citet{Draine+Lazarian_1998a,Draine+Lazarian_1998b} and suggested that this discrepancy was
due to PAH depletion in the WIM. However, subsequent
studies have suggested an AME amplitude a factor of $\sim3$ lower than
the earlier estimates even in Galactic clouds \citep{Tibbs+etal_2010,
  Tibbs+etal_2011, Planck_Int_XV, Planck_Int_XVII}. Thus, the
results of \citet{Dobler+Draine+Finkbeiner_2009} indicate instead a robustness
of the AME strength across environments and, together with our
results, further evidence that the AME is not associated
with PAHs.

The correlation between $f_{\rm PAH}$ and $\tau_{353}$ also sheds
light on the apparent negative correlation observed between $f_{\rm
  PAH}$ and $I_\nu^{\rm AME}/\tau_{353}$ in
Figure~\ref{fig:residual}a. Figures~\ref{fig:corr}a and \ref{fig:mbb_corr}c demonstrate that $\alpha\tau_{353}$ tends to
underpredict the true AME intensity at low values of $\tau_{353}$ and to
overpredict at high values. Since $f_{\rm PAH}$ is positively
correlated with $\tau_{353}$, it is not surprising that assuming a
linear relationship between the AME and $f_{\rm
  PAH}\tau_{353}$ only exacerbates those discrepancies. 

\subsection{Correlation with the Radiation Field}

\begin{figure*}
    \centering
        \scalebox{1.0}{\includegraphics{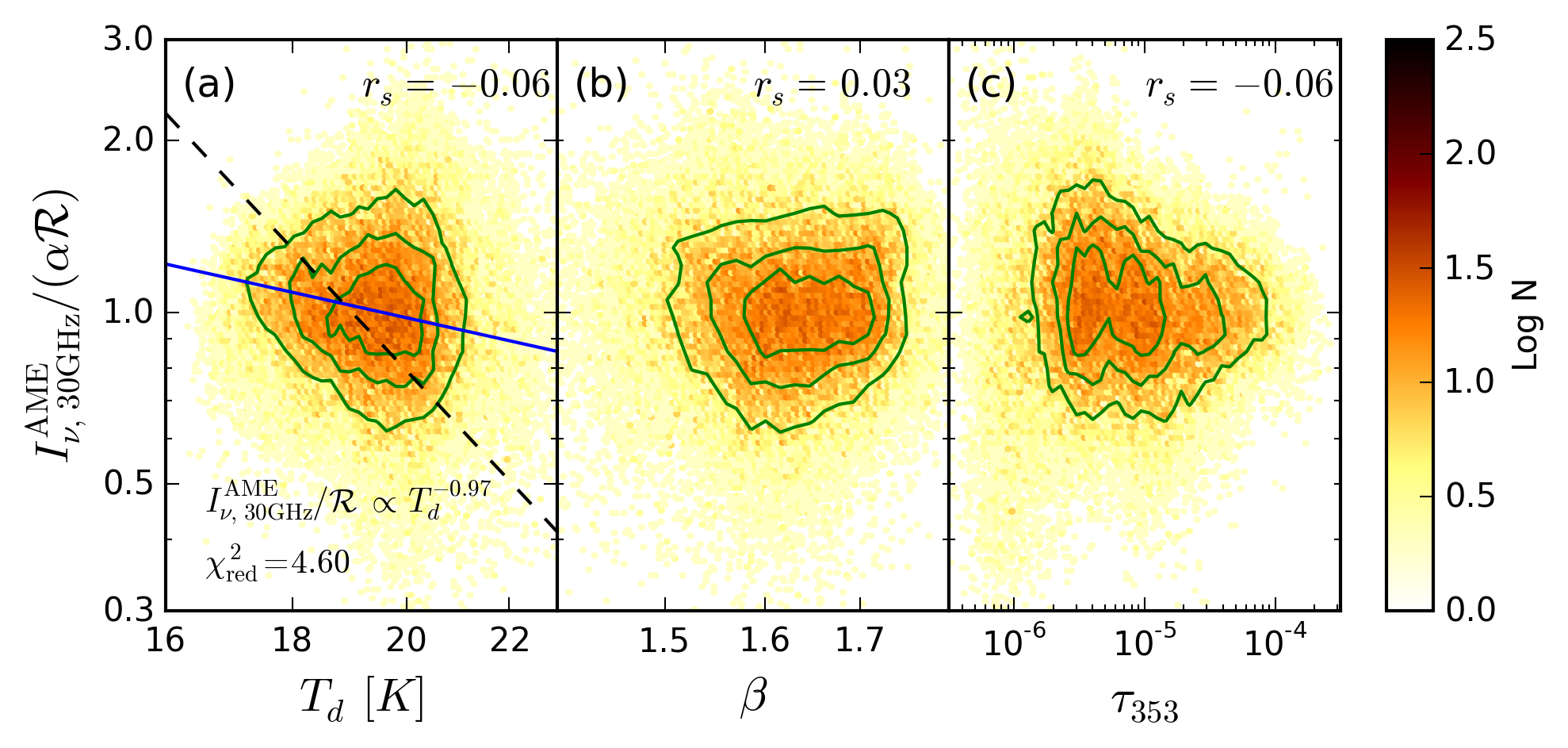}}  
    \caption[Dependence of AME Intensity Upon Temperature, $\beta$,
    and $\tau_{353}$]{The 30\,GHz AME intensity normalized by
      $\alpha\mathcal{R}$ is plotted against (a) $T_d$ and (b) $\beta$ determined by
      modified blackbody fits to the dust SED \citep{Planck_2013_XI}, and
      against $\tau_{353}$ determined from component separation
      \citep{Planck_2015_X} in panel (c). Since PAHs are
      depleted in dense regions, the spinning PAH hypothesis predicts
      that the AME per $\mathcal{R}$ should be smaller in denser
      regions. These regions are likely to have more cold dust and
      thus a smaller $\beta$, but no correlation is observed
      between $\beta$ or $\tau_{353}$ and the AME intensity per
      $\mathcal{R}$. The data do suggest a possible correlation with
      $T_d$. We find a best-fit power-law of $T_d^{-0.97}$ (blue solid) and
      plot also the best fit relation assuming the AME is thermal emission with $\tau_{30}\propto\tau_{353}$, i.e.
        $I_\nu^{\rm AME}\propto\tau_{353}T_d$, hence $I_\nu^{\rm AME}/R\propto
        T_d^{-4.65}$ (black dashed). The latter
      relationship is disfavored relative to the former
      ($\chi^2_{\rm red} = 10.2$ vs. 4.6). Finally, we note
      that our conclusions are unchanged when we redo this analysis
      using $T_d$ and $\beta$ determined by \citet{Planck_2015_X}.} \label{fig:mbb_corr} 
\end{figure*}

\begin{figure*}
    \centering
        \scalebox{0.8}{\includegraphics{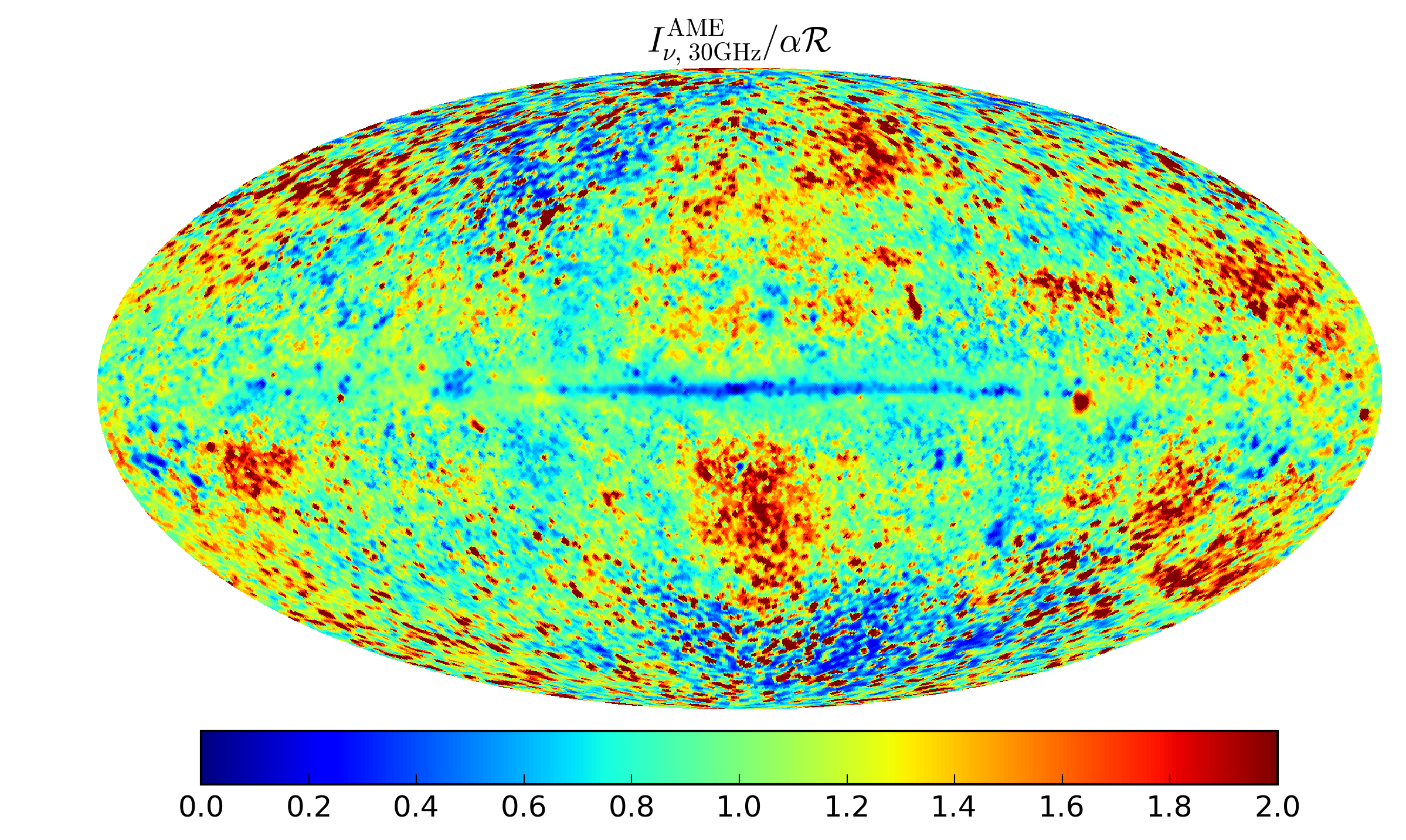}}  
    \caption[AME/R]{The Mollweide projection full-sky map of the
      30\,GHz AME intensity normalized by $\alpha\mathcal{R}$. Although
      some large-scale features associated with strong synchrotron
      emission are present, overall the map has little correlation
      with the synchrotron intensity ($r_s = 0.11$).} \label{fig:ame_r_map} 
\end{figure*}

Figure~\ref{fig:corr} and Table~\ref{table:corr} also indicate the
surprising result that the AME is more tightly correlated with
$\mathcal{R}$ than $\tau_{353}$. We would expect $\mathcal{R}$ and
$\tau_{353}$ to be related through the strength of the radiation
field-- a fixed quantity of dust will radiate more when exposed to
more radiation. The most straightforward conclusion
is that the AME is enhanced by a stronger radiation field,
which runs counter to the predictions of the spinning dust
hypothesis \citep{Ali-Haimoud+Hirata+Dickinson_2009,
  Ysard+Verstraete_2010}.

A positive correlation between the radiation field strength and the AME
intensity has been noted in both the Perseus molecular cloud
\citep{Tibbs+etal_2011} and the H{\sc II} region RCW175
\citep{Tibbs+etal_2012}. This trend was also noted by
\citet{Planck_Int_XV}, who attribute the
correlation to a positive correlation between the radiation field and
the local gas density.

The relationship between the AME/$\mathcal{R}$ and $T_d$, illustrated in
\ref{fig:mbb_corr}a, is
best-fit by a power-law of index -0.97. If the AME were thermal
emission in the Rayleigh-Jeans limit, we would expect $I_\nu^{\rm
  AME}$ to scale linearly
with $T_d$, i.e $I_{\nu,\ 30\,{\rm GHz}}^{\rm AME} \propto \tau_{30}T_d$. Since $\mathcal{R} \propto T_d^{4+\beta}$ if $\tau_\nu
\propto \nu^\beta$, $I_\nu^{\rm
  AME}/\mathcal{R}$ should therefore scale as $T_d^{-\beta - 3} \approx
T_d^{-4.65}$ if the AME is thermal emission and $\beta \approx 1.65$. Fitting the data with a power $T_d^{-4.65}$ yields a substantially worse
$\chi^2_{\rm red}$ (10.2 vs. 4.6). The
data suggest then that the AME is nonthermal emission, unless the dust
opacity at 30\,GHz has a steep temperature dependence
(e.g. $\tau_{30}/\tau_{353} \propto T_d^{3.7}$, where $\tau_{30}$ is
the dust optical depth at 30\,GHz.). We caution that systematic effects
arising from fitting a single-$T_d$ modified blackbody to the thermal
dust emission may also alter the correlation with $T_d$ and we thus
cannot rule out a thermal emission mechanism completely based upon
these data alone.

If the fit AME component is contaminated with emission from other low-frequency
foregrounds, then this may also induce correlations with
$T_d$. Indeed, Figure~\ref{fig:ame_r_map} reveals some coherent
large-scale structures in the map of AME/$\mathcal{R}$ that are likely
related to strong synchrotron emission
\citep[see][Figure~46]{Planck_2015_X} and perhaps the {\it Fermi}
bubbles \citep{Su+Slatyer+Finkbeiner_2010}. We
discuss this possibility further in Section~\ref{sec:discussion},
though aside from a few clear structures, contamination appears to be
minimal.

For pixels containing regions with both high and low radiation
intensities, fitting the $\lambda \geq 100\,\mu$m emission by a
single-temperature modified blackbody will lead to systematic errors,
tending to overestimate $T_d$ and underestimate both $\beta$ and
$\tau_{353}$. Thus, if the correlation between the AME and the dust
radiance is driven by the correlation between the radiation field
intensity and the local gas density, we might expect the AME per
$\mathcal{R}$ to correlate with $\beta$. We find no evidence for such a
correlation (see Figure~\ref{fig:mbb_corr}b).

\subsection{Correlation with $I_\nu^{100\,\mu{\rm m}}$}
The ratio of $I_{\nu,\ 30\,{\rm GHz}}^{\rm AME}$ to $I_\nu^{100\,\mu{\rm m}}$ is often
quoted in the literature. We obtain a value of
$(2.88\pm0.65)\times10^{-4}$ (see Table~\ref{table:corr}), consistent with other determinations. For
instance, performing component separation on WMAP observations at
intermediate Galactic latitudes, \citet{Davies+etal_2006} derived a
ratio of $3\times10^{-4}$. \citet{Alves+etal_2010} likewise find a
ratio of $3\times10^{-4}$ combining observations of radio
recombination lines in the Galactic plane with WMAP data. Studying
H{\sc II} regions in the Galactic plane with the Very Small Array at
33 GHz, \citet{Todorovic+etal_2010} find a ratio of
$1\times10^{-4}$. Analyzing a sample of 98 Galactic clouds with {\it
  Planck} observations, \citet{Planck_Int_XV} derived a ratio of
$2.5\times10^{-4}$. Thus, the AME component identified by
\citet{Planck_2015_X} has a strength relative to the 100\,$\mu$m dust emission
in good agreement with what has been observed in other studies.

However, as discussed in detail
by \citet{Tibbs+Paladini+Dickinson_2012}, this ratio is subject to
significant variability. While the 100\,$\mu$m emission
is a reasonable proxy for the total dust luminosity, its sensitivity
to the dust temperature introduces significant non-linearities in the
relationship. This effect is evident at particularly low and high
values of $I_\nu^{100\,\mu{\rm m}}$ in
Figure~\ref{fig:corr}d as evidenced by the somewhat non-linear shape of the
scatter plot. As expected, the tightness of the correlation
between the AME and 100\,$\mu$m increases when the analysis is
restricted to a set of pixels with similar $T_d$.

\subsection{Correlation with $I_\nu^{12\,\mu{\rm m}}$}
Finally, we find that the 12\,$\mu$m emission is also tightly
correlated with the AME with dispersion only slightly less than that of
$\mathcal{R}$. While it is tempting to read this as a vindication of
the spinning PAH model, the foregoing analysis suggests that this
tight correlation is merely the product of the 12\,$\mu$m emission
being an excellent tracer of both the dust column and the radiation
field strength. $f_{\rm PAH}$, i.e. the 12\,$\mu$m emission per unit
$\mathcal{R}$, does not correlate with the AME intensity (see Figures~\ref{fig:residual}b
and~\ref{fig:mask_corr}). 

\begin{figure}
    \centering
        \scalebox{1.0}{\includegraphics{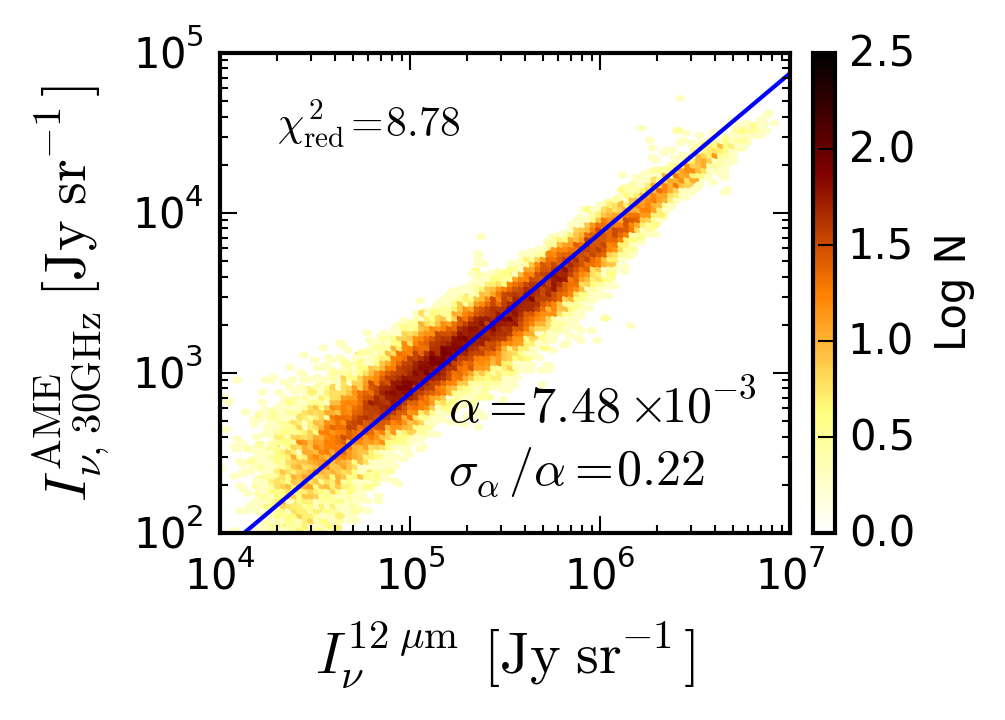}}  
    \caption[Correlation Between $I_\nu^{\rm AME}$ and
    $I_\nu^{12\,\mu{\rm m}}$]{As in each panel of Figure~\ref{fig:corr}, we plot the
      WISE 12\,$\mu$m intensity against the 30\,GHz AME intensity. The
    correlation is comparably tight as observed with $\mathcal{R}$.} \label{fig:wise_corr} 
\end{figure}

Thus, while the PAH emission is an excellent predictor of the AME strength, this
appears to be by virtue of being an excellent predictor of the dust
radiance rather than the result of an inherent link between the AME
and PAHs.

\subsection{Emission from Magnetic Dust}
If the AME is not spinning PAHs, could it be emission from magnetic
grains? \citet{Draine+Lazarian_1999} predicted that, unlike spinning
PAH emission, magnetic dust emission would be equally strong per dust
mass in both dense and diffuse regions. Since the dust in dense
regions will be cooler than that in diffuse regions, pixels with
significant dust emission from both diffuse and dense regions will
have broader SEDs and thus are fitted by smaller
values of $\beta$ relative to diffuse regions when fitting the SED
with a modified single $T_d$
blackbody. Thus, the AME per dust mass is predicted to be {\it negatively}
correlated with $\beta$ in the spinning PAH model and uncorrelated in
the magnetic dust model. In Figure~\ref{fig:mbb_corr}b we demonstrate
that AME/$\mathcal{R}$ and $\beta$ are largely uncorrelated ($r_s = 0.03$). However, magnetic grains should emit thermally
\citep{Draine+Hensley_2013}, whereas the relationship
derived above between $I_\nu^{\rm AME}/\mathcal{R}$ and
$T_d$ favors a non-thermal emission mechanism. In addition, thermal
emission from magnetic dust seems likely to be polarized
\citep{Draine+Hensley_2013}  whereas observations find the AME to be
minimally polarized (see Section~\ref{sec:discussion} for further
discussion of polarization observations).

\subsection{Dependence on Masks}
\label{subsec:masks}
\begin{figure*}
    \centering
        \scalebox{1.0}{\includegraphics{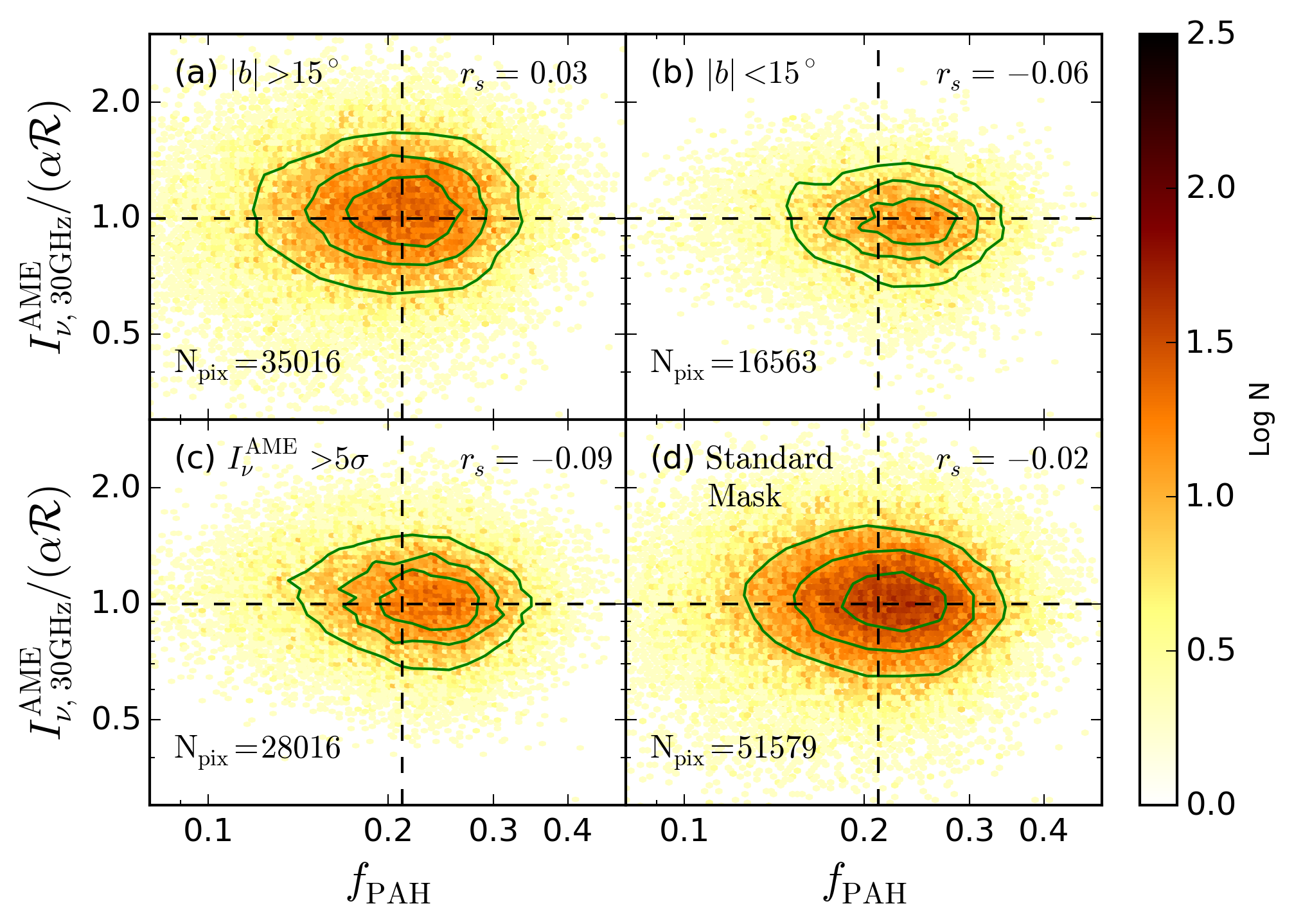}}  
    \caption[Dependence of Correlation Between $f_{\rm PAH}$ and
    $I_\nu^{\rm AME}$ Upon Masks]{As in Figure~\ref{fig:residual}, $f_{\rm PAH}$ is plotted
    against the 30\,GHz AME intensity normalized by
    $\alpha\mathcal{R}$. We examine four different masks: (a) considering only higher
  Galactic latitudes $|b| > 15^\circ$, (b) considering only lower
  Galactic latitudes $|b| < 15^\circ$, (c) considering only pixels
  in which at least one AME component was significant at $>5\sigma$,
  and (d) using the standard mask only. For reference, the dashed
  lines mark the median values in the standard mask. In all cases, there is no
  evidence for positive correlation between the AME and $f_{\rm PAH}$.} \label{fig:mask_corr} 
\end{figure*}

The lack of correlation between $f_{\rm PAH}$ and the AME intensity
per $\tau_{353}$ or $\mathcal{R}$ is a potentially serious problem for the spinning
PAH hypothesis. We thus test the sensitivity of this result to the
region of the Galaxy examined. In Figure~\ref{fig:mask_corr}, we
perform the same analysis as in Figure~\ref{fig:residual}b and quantify the degree of
correlation with the Spearman correlation coefficient $r_s$.

Starting with the $N = 51579$ pixels remaining unmasked following the cuts
discussed in~\ref{subsec:galmask}, in Figure~\ref{fig:mask_corr}a we consider only the pixels with
Galactic latitude $|b| > 15^\circ$. The behavior very much mimics
that observed in Figure~\ref{fig:residual}b, with no compelling
evidence of a correlation. In Figure~\ref{fig:mask_corr}b, we examine
pixels with $|b| < 15^\circ$. These pixels have
more PAH emission per dust radiance (i.e. higher values of $f_{\rm PAH}$) than
those at higher latitudes, as would be expected from PAH destruction
in diffuse regions. Again,
however, there is no indication of a correlation of the AME intensity
per $\mathcal{R}$ with $f_{\rm PAH}$.

We next examine in
Figure~\ref{fig:mask_corr}c only
those pixels in which one or both of the AME components are
significant at greater than $5\sigma$. The cut on AME significance
does not change significantly the behavior observed in
Figure~\ref{fig:mask_corr}a other than to eliminate some pixels with low
$f_{\rm PAH}$ values. As illustrated in Figure~\ref{fig:fpah_tau_r}, the pixels
with low $f_{\rm PAH}$ also tend to have low surface
brightness, which may be responsible for inducing the slight negative
correlation observed. Finally, Figure~\ref{fig:mask_corr}d shows for comparison the same
analysis performed on the standard mask (Figure~\ref{fig:residual}a).

The non-correlation of the AME intensity and $f_{\rm
  PAH}$ is therefore robust to assumptions either on the AME significance or the
region of the sky analyzed. We now turn to the implications of this
result in the following section.

\section{Discussion}
\label{sec:discussion}

\begin{figure*}
    \centering
        \scalebox{1.0}{\includegraphics{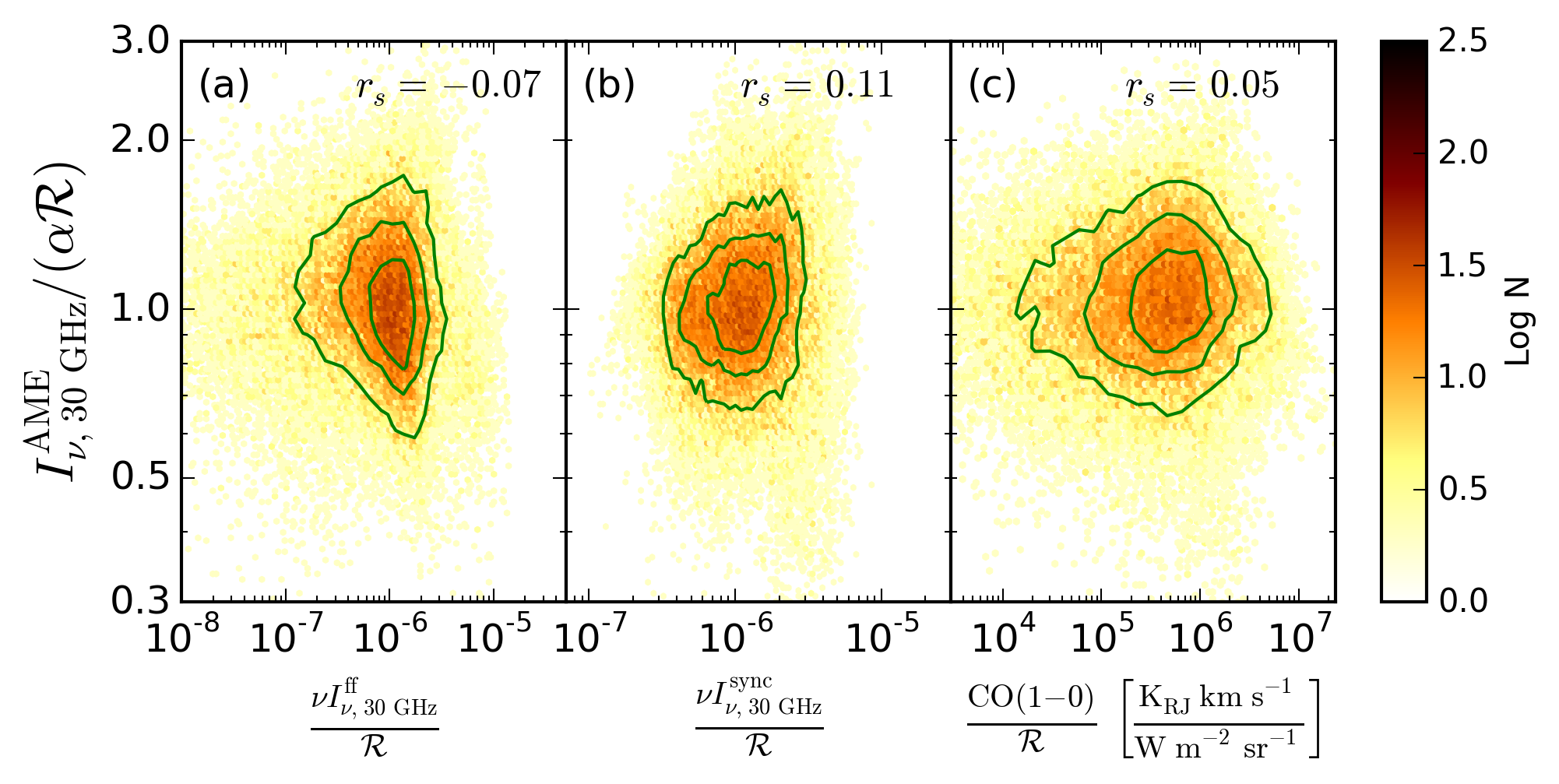}}  
    \caption[Correlation Between the AME and Other Low-Frequency
    Foregrounds]{The 30\,GHz AME intensity normalized by
      $\alpha\mathcal{R}$ is plotted against (a) the 30\,GHz free-free
      intensity, (b) the 30\,GHz synchrotron intensity, and (c) the
      CO(1-0) line emission all normalized by $\mathcal{R}$. There is
      evidence for weak correlation in all panels.} \label{fig:low_freq}
\end{figure*}

\begin{figure*}
    \centering
        \scalebox{1.0}{\includegraphics{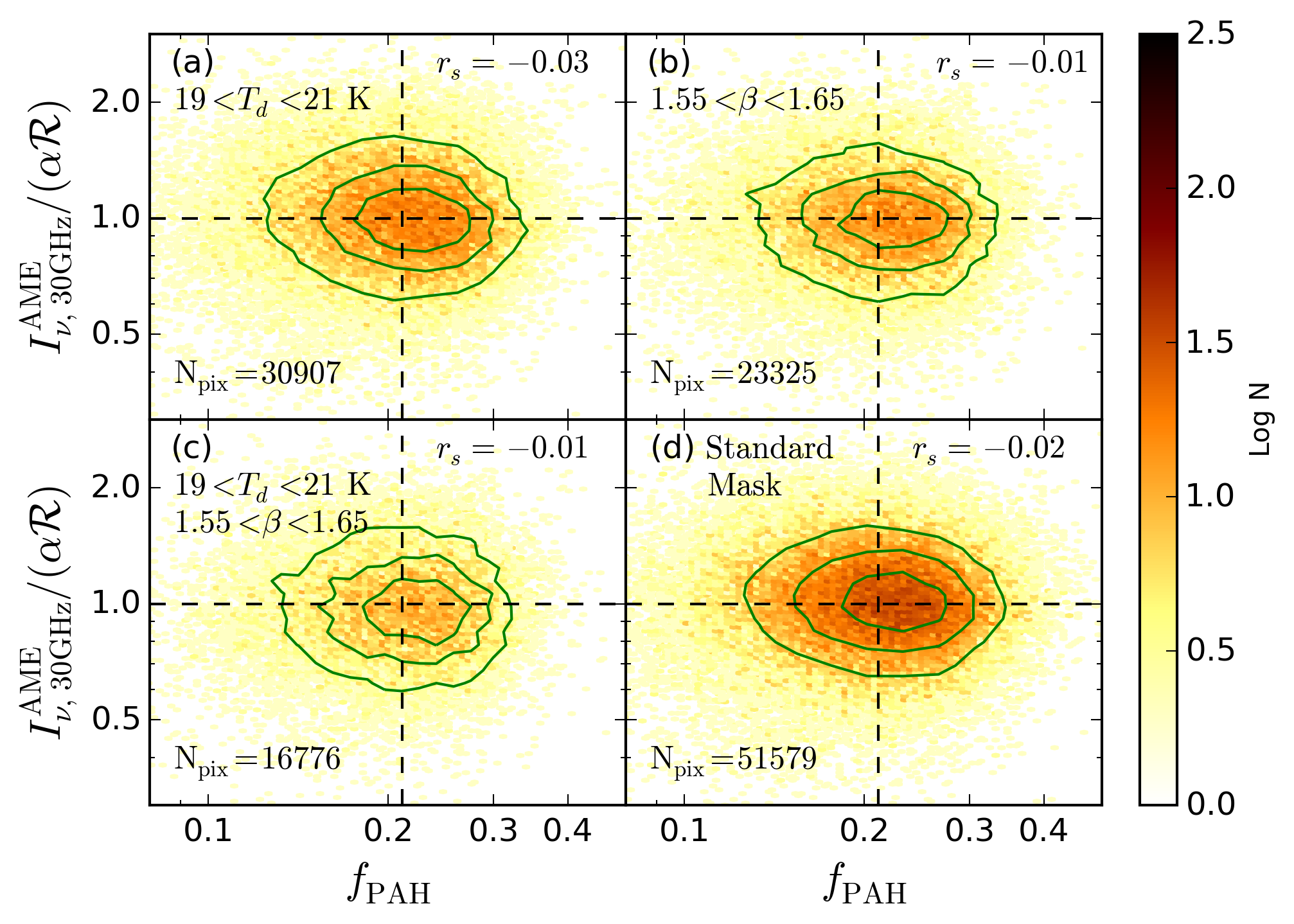}}  
    \caption[Dependence of Correlation Between $f_{\rm PAH}$ and
    $I_\nu^{\rm AME}$ Upon Environment]{As in Figure~\ref{fig:mask_corr}, $f_{\rm PAH}$ is plotted
    against the 30\,GHz AME intensity normalized by
    $\alpha\mathcal{R}$. We examine four different sets of pixels: (a)
    pixels with $T_d$ between 19 and 21\,K (typical of the
    high-latitude sky), (b) pixels with $\beta$ between 1.55 and 1.65,
    (c) pixels meeting both the criteria of panels (a) and (b),
  and (d) using the standard mask only. For reference, the dashed
  lines mark the median values in the standard mask. Thus even when
  the conditions of the local environment are held relatively fixed, there is no
  evidence for a positive correlation between the AME and $f_{\rm PAH}$.} \label{fig:tcut} 
\end{figure*}

One of the largest sources of uncertainty in this analysis is the
fidelity of the AME spectrum recovered from decomposition. Due to a
lack of data between the WMAP 23\,GHz band and the Haslam data at
408\,MHz, it is difficult to constrain the relative contributions of
the AME, synchrotron, and free-free in the frequency range of
interest. \citet{Planck_2015_X} notes that their estimates of
synchrotron emission are lower than the 9\,yr. WMAP analysis \citep{Bennett+etal_2013} by about
70\% at high Galactic latitudes and factors of several in the Galactic
plane, with the
AME component estimate being larger as a result. It is therefore possible that
the AME map has a non-negligible synchrotron component.

The presence of contamination in the AME map, such as free-free or
synchrotron, would of course mean that $f_{\rm PAH}$ is not related to
the inferred AME intensity in a perfectly linear way even if the AME comes from
spinning PAHs. Further, since $f_{\rm PAH}$ correlates with
$\tau_{353}$ and $\mathcal{R}$, it is also expected to
correlate with the free-free and synchrotron emission, which
increase with increasing gas (and therefore dust) column density.

As illustrated in Figure~\ref{fig:low_freq}a, $I_\nu^{\rm
  AME}/\alpha\mathcal{R}$ is weakly anti-correlated with the 30\,GHz free-free emission per
$\mathcal{R}$ ($r_s = -0.07$).

In Figure~\ref{fig:low_freq}b, we plot $I_\nu^{\rm
  AME}/\alpha\mathcal{R}$ against the 30\,GHz synchrotron intensity per
$\mathcal{R}$. The two
quantities show evidence of a weak positive correlation ($r_s =
0.11$).

In Figure~\ref{fig:low_freq}c, we plot $I_\nu^{\rm
  AME}/\alpha\mathcal{R}$ against the CO(1-0) line intensity per
$\mathcal{R}$. The two
quantities show possible evidence of a weak positive correlation ($r_s =
0.05$).

It is likely that some degree of contamination from other
low-frequency components exists in the AME maps. This is to be expected given the theoretical difficulties
modeling their spectra and the relative lack of observational
constraints at very low frequencies. In the future, more detailed decomposition enabled by additional data will
improve the fidelity of the inferred AME spectrum and test the
conclusions derived here. We note that the observed correlations
between the AME/$\mathcal{R}$ and the other low frequency components are
rather weak, making it unlikely that these components
are driving the observed non-correlation between the AME and $f_{\rm PAH}$.

Even assuming no contamination in the fit AME component, correlations
with the synchrotron emission are plausible. For instance, if the AME
arises from spinning ultrasmall grains, it might be affected in
synchrotron-bright supernova remnants where it may be enhanced if
shattering in grain-grain collisions increases the ultrasmall grain
population, or suppressed if ultrasmall grains are destroyed by sputtering.

The theoretical uncertainties in the models for these
emission components underscore the need for obtaining ancillary data
at lower frequency. Upcoming 5\,GHz observations from the C-Band All-Sky Survey
(C-BASS) \citep{King+etal_2014} and 2.3\,GHz observations from the S-Band
Polarized All-Sky Survey (S-PASS) \citep{Carretti+etal_2009} will play an invaluable role in
disentangling the low-frequency components.

A second source of uncertainty is the ability of $f_{\rm PAH}
\equiv \Delta\left(\nu I_\nu\right)^{12\,\mu{\rm m}}/\Delta\mathcal{R}$ to trace
the PAH abundance. In particular, the fraction of the PAH emission
appearing in the WISE 12\,$\mu$m band
can depend on the ionization state of the PAHs and other properties of
the local environment \citep[e.g.][]{Draine+Li_2007, Draine_2011b}. Thus it
may be possible to ``wash out'' a correlation between $f_{\rm PAH}$
and the AME intensity even if the AME arises from spinning PAHs. 

To test the plausibility of such a scenario, in Figure~\ref{fig:tcut} we correlate the AME
intensity per $\mathcal{R}$ with $f_{\rm PAH}$ only in pixels with
similar environmental conditions as determined by their fit dust
temperature and $\beta$. We find no significant differences from our
primary analysis and thus no evidence that variations in the local
environment are driving the lack of correlation between $f_{\rm PAH}$
and the AME intensity per $\mathcal{R}$.

In addition to the uncertainties discussed above, each observable $A$ also has an associated
observational uncertainty $\sigma_A$ which we have not included in
Equation~\ref{eq:likelihood}. $\sigma_A$ is highly degenerate with
$\sigma_\alpha$ and thus it is difficult to quantify their relative
contributions to the total uncertainty. Instead, we ask what assumed
fractional uncertainty on $A$ would be required for the data to be
consistent with an intrinsic relationship with $\sigma_\alpha/\alpha =
0.2$, comparably tight as found with
$\mathcal{R}$. We denote this quantity $\sigma_A/A\left(0.2\right)$ in
Table~\ref{table:corr}, finding that for $\tau_{353}$ and $f_{\rm
  PAH}\tau_{353}$, $\sigma_A/A$ would
need to be larger than 25\% for the observations to be consistent with
an intrinsic relationship with so narrow a dispersion. The small
uncertainties on the thermal dust emission at 353\,GHz reported by \citep{Planck_2015_X}
are inconsistent with what would be required for the data to be
compatible with a tight intrinsic relationship between AME and either
$\tau_{353}$ or $f_{\rm PAH}\tau_{353}$. Thus, it does not seem plausible to attribute the
entirety of our findings to the relative uncertainties of the
observables.

Finally, \citet{Planck_2015_XXV} demonstrate that the correlation
between AME/$\mathcal{R}$ and $f_{\rm PAH}$ can be affected by
systematic errors in the determination of $\mathcal{R}$. Although an
analysis using the \texttt{Commander}-derived $\mathcal{R}$ and 
$f_{\rm PAH}$ determined by straight division of the $W3$ and
$\mathcal{R}$ maps is inappropriate for the reasons discussed in
Section~\ref{subsec:data_wise}, it remains possible that the $\mathcal{R}$
map produced by \citet{Planck_2013_XI} is biased in ways that affect the
correlation with $f_{\rm PAH}$. Dust modeling employing
additional data, such as 60\,$\mu$m dust emission, will help clarify
this issue. In the meantime, we note that $\mathcal{R}$, being the
integral of the FIR dust intensity, is rather insensitive to the
specifics of modeling as long as the model provides a reasonable fit
to the data. Thus, particularly when 100\,$\mu$m data is employed,
$\mathcal{R}$ is straightforward to determine and we have no reason to
suspect strong systematic biases.

If indeed the AME is {\it not} correlated with the PAH abundance
and {\it is}
correlated with the strength of the radiation field, what are the
implications for the origin of the emission? We present two
possibilities:

\begin{enumerate}
\item The AME is spinning dust emission that arises primarily from ultrasmall 
grains that are not PAHs. \citet{Li+Draine_2001} have shown that as much as
$\sim10$\% of the interstellar silicate mass could be in ultrasmall
grains. However, it remains to be seen whether these grains could
produce enough rotational emission to account for the AME.

\item The AME is predominantly thermal dust emission, such as magnetic dipole
emission from magnetic materials. However, the relationship between
AME/$\mathcal{R}$ and $T_d$ does not appear consistent with thermal
emission (see Figure~\ref{fig:mbb_corr}). Additionally, current models of magnetic dipole emission do not predict
behavior that could emulate the observed AME SED without
invoking highly elongated (e.g., 5:1 prolate spheroids) Fe inclusions
\citep{Draine+Hensley_2013}. Furthermore, thermal dust emission from
an aligned component of
the grain population is expected to be significantly polarized, 
including magnetic dipole emission from ferromagnetic
inclusions \citep{Draine+Hensley_2013}, at odds with current
non-detections of AME polarization. 
Nevertheless, the theoretical calculations of the emissivities of these magnetic materials
are still quite uncertain, and more
laboratory data is needed to assess the behavior of magnetic
materials at microwave frequencies to determine whether such grains
could be a potential source of the AME. It remains conceivable that
some interstellar grain component might produce the AME by
non-rotational electric dipole radiation, but we are not aware of
materials that could do this.
\end{enumerate}

Invoking an alternative explanation for the AME also requires
explaining why the PAHs are {\it not} a substantial source of spinning
dust emission. Because the PAHs are clearly present and must be
rotating, this would require that the electric dipole
moments of the PAHs have been significantly overestimated (the spinning
dust emission scales as the square of the dipole
moment). The electric dipole moments of selected hydrocarbon molecules compiled by
\citet{Draine+Lazarian_1998b} have a scatter of nearly an order of
magnitude. Further, harsh UV irradiation more easily destroys asymmetric
molecules, perhaps preferentially selecting for a population of more
symmetric PAHs with smaller dipole moments. Thus it is
plausible that the dipole moment distributions adopted in spinning
dust models may significantly overestimate the true electric dipole
moments of interstellar PAHs.

\citet{Lazarian+Draine_2000} estimated the polarization fraction of spinning dust
emission to be $p \lesssim 0.01$ near 30\,GHz. This estimate would
also apply to spinning dust emission from non-PAH grains. However, if the AME is thermal emission from large aligned
grains, it should be significantly polarized with ${\bf
  E} \perp {\bf B}_0$ for electric dipole radiation or ${\bf
  E} \parallel {\bf B}_0$ for magnetic dipole radiation from magnetic
inclusions \citep{Draine+Hensley_2013}. 

Observations of known AME sources in polarization suggest minimal
polarization of the AME. In the Perseus molecular cloud, the
polarization fraction of the total emission was found to be $3.4^{+1.5}_{-1.9}$\% at 11\,GHz
\citep{Battistelli+etal_2006}, less
than 6.3\% at 12\,GHz \citep{GenovaSantos+etal_2015}, less
than 2.8\% at 18\,GHz \citep{GenovaSantos+etal_2015}, and less
than $\sim 1$\% at 23\,GHz \citep{LopezCaraballo+etal_2011,
  Dickinson+Peel+Vidal_2011}. Likewise, observations of the
$\rho$ Ophiuchi molecular cloud have yielded upper limits of
$\lesssim1$\% at 30\,GHz
\citep{Casassus+etal_2008,Dickinson+Peel+Vidal_2011}. \citet{Mason+etal_2009}
placed an upper
limit of 2.7\% on the 9.65\,GHz polarization fraction in the dark cloud
Lynds 1622. 21.5\,GHz observations of the H{\sc ii} region RCW175
yielded a polarization fraction of $2.2\pm0.4$\%
\citep{Battistelli+etal_2015}, though it is unclear whether this
polarization is arising from the AME or a sub-dominant synchrotron
component.

These upper limits on polarization in
the 10 - 30\,GHz emission appear to favor spinning dust emission from a
non-PAH population of ultrasmall grains.

\section{Conclusion}
\label{sec:conclusions}
We have combined the {\it Planck} foreground component maps, {\it
  Planck} modified blackbody dust parameter maps, and WISE 12\,$\mu$m
maps to test key predictions of the spinning PAH hypothesis. The
principal conclusions of this work are as follows:

\begin{enumerate}
\item $\tau_{353}$, the dust radiance $\mathcal{R}$, and
  $I_\nu^{12\,\mu{\rm m}}$ are all excellent predictors of the
  30\,GHz AME intensity. $\mathcal{R}$
  exhibits the tightest correlation,
  suggesting that the AME is sensitive to the strength of the
  radiation field.
\item Neither AME/$\tau_{353}$ nor AME/$\mathcal{R}$ show any correlation with the PAH
  emission whether considering the full sky, regions close to the
  Galactic plane, or higher Galactic latitudes.
\item We find that $f_{\rm PAH}$ is correlated with both
  $\tau_{353}$ and $\mathcal{R}$, consistent with PAH destruction in
  low density regions.
\item Taken together, these facts pose a serious challenge to the
  spinning PAH paradigm as the explanation for the AME. Alternative
  explanations, such as magnetic dipole emission from ferro- or
  ferrimagnetic grains, should be more thoroughly investigated.
\item More low
  frequency constraints are needed to break degeneracies between the
  AME, free-free, and synchrotron to enable more accurate decomposition and
  to better constrain the AME spectrum. Upcoming all-sky observations
  from C-BASS and S-PASS will thus facilitate deeper investigations into
  the origin of the AME.
\item Further measurements of AME polarization will help clarify the nature of
  the grains responsible for the AME.
\end{enumerate}

\acknowledgments
{We thank Kieran Cleary, Hans
  Kristian Eriksen, Doug Finkbeiner,
  Chelsea Huang, Alex Lazarian, Mike Peel, David Spergel, Ingunn
  Wehus, and Chris
  White for stimulating
  conversations. BSH and BTD acknowledge support from NSF grant
  AST-1408723. The research was carried out in part at the Jet Propulsion
  Laboratory, California Institute of Technology, under a contract
  with the National Aeronautics and Space Administration. This work was supported in part by the Director, Office of Science,
Office of High Energy Physics, of the U.S. Department of Energy under
contract No. DE-AC02-05CH11231.}

\bibliography{planck_ame}

\end{document}